\documentclass[]{article}

\usepackage[margin=1.0in]{geometry}
\usepackage[capposition=top]{floatrow}
\usepackage{amsmath}
\usepackage{amsfonts}
\usepackage{graphicx}

\title{Transportability without positivity: a synthesis of statistical and simulation modeling}
\author{Paul N Zivich\textsuperscript{1,2}, 
	    Jessie K Edwards\textsuperscript{2}, 
	    Eric T Lofgren\textsuperscript{3}, 
	    Stephen R Cole\textsuperscript{2}, \\
	    Bonnie E Shook-Sa\textsuperscript{4}, 
	    Justin Lessler\textsuperscript{2,5,6}}
\date{%
	\small
	\textsuperscript{1}Institute of Global Health and Infectious Diseases, University of North Carolina at Chapel Hill, Chapel Hill, NC\\%
	\textsuperscript{2}Department of Epidemiology, Gillings School of Global Public Health, University of North Carolina at Chapel Hill, Chapel Hill, NC\\%
	\textsuperscript{3}Paul G. Allen School for Global Health, Washington State University, Pullman, WA\\%
	\textsuperscript{4}Department of Biostatistics, Gillings School of Global Public Health, University of North Carolina at Chapel Hill, Chapel Hill, NC\\%
	\textsuperscript{5}Carolina Population Center, University of North Carolina at Chapel Hill, Chapel Hill, NC\\[2ex]%
	\textsuperscript{6}Department of Epidemiology, Johns Hopkins Bloomberg School of Public Health, Baltimore, MD\\[2ex]%
	\today
}

\begin{document}

\maketitle

\begin{abstract}
	When estimating an effect of an action with a randomized or observational study, that study is often not a random sample of the desired target population. Instead, estimates from that study can be transported to the target population. However, transportability methods generally rely on a positivity assumption, such that all relevant covariate patterns in the target population are also observed in the study sample. Strict eligibility criteria, particularly in the context of randomized trials, may lead to violations of this assumption. Two common approaches to address positivity violations are restricting the target population and restricting the relevant covariate set. As neither of these restrictions are ideal, we instead propose a synthesis of statistical and simulation models to address positivity violations. We propose corresponding g-computation and inverse probability weighting estimators. The restriction and synthesis approaches to addressing positivity violations are contrasted with a simulation experiment and an illustrative example in the context of sexually transmitted infection testing uptake. In both cases, the proposed synthesis approach accurately addressed the original research question when paired with a thoughtfully selected simulation model. Neither of the restriction approaches were able to accurately address the motivating question. As public health decisions must often be made with imperfect target population information, model synthesis is a viable approach given a combination of empirical data and external information based on the best available knowledge.
\end{abstract}

\section*{Introduction}
To aid public health decision making, epidemiologists produce quantitative effect estimates of different actions (e.g., exposure, treatments, interventions) for specific populations. However, data are often not a random sample from the desired target population, a problem shared by both randomized trials and observational studies. Instead, one can attempt to `transport' estimates from a separate study sample to the target population \cite{westreich_transportability_2017, dahabreh_generalizing_2019, dahabreh_toward_2020, bareinboim_causal_2016, degtiar_review_2023, keiding_perils_2016, cerda_systems_2019}. Methods for transportability have generally proceeded under the assumption that population membership and potential outcomes are independent conditional on a set of observed covariates (i.e., conditional exchangeability) \cite{westreich_transportability_2017, dahabreh_generalizing_2019, dahabreh_toward_2020, degtiar_review_2023}. This assumption requires that all relevant covariate patterns in the target population are also observed in the study population (i.e., deterministic positivity \cite{zivich_positivity_2022}). However, strict eligibility criteria or practical constraints may lead to positivity violations.

Here, we consider transportability when positivity does not hold. We review two common approaches to addressing positivity violations: restricting the target population to regions where the positivity assumption holds, and restricting the adjustment set of covariates to a subset where the positivity assumption holds. Both approaches have problems: restricting the target population no longer addresses the motivating question, and restricting the adjustment set requires that the excluded covariates induce little-to-no bias. To avoid these shortcomings, we propose an alternative that uses a synthesis of statistical and simulation modeling. Each approaches is applied to an illustrative example on improving sexually transmitted infection (STI) testing coverage (i.e., proportion tested).

\section*{Motivating Problem}

Consider the following scenario: a clinical colleague approaches us with the goal of increasing STI testing coverage among their patients. They wish to offer patients either electronically ordered self-sampling STI test kits (e-STI) or standard walk-in STI testing via text message. While the clinic is unable to conduct a trial comparing testing modalities, our colleague has access to data from a randomized trial comparing these strategies at a different clinic and data on key demographics of their clinic's patients. 

Let $A_i$ indicate the STI testing information presented via text message to patient $i$, with $A_i=1$ being e-STI and $A_i=0$ being walk-in STI testing. Let $Y_i^a$ indicate the potential outcome of STI testing uptake under the STI testing option $a$ in the following 6-weeks, and $Y_i$ be the observed outcome. Additionally, let $V_i$ indicate age (years) and $W_i$ indicate gender (male, female). Finally, $S_i$ indicates whether patient $i$ was in the clinic data ($S_i=1$) or the randomized trial data ($S_i=0$). For inference, we proceed under a population model, where data are considered to be a random sample of a much larger population \cite{lehmann_elements_2004}.

Our colleague's question is: what is the difference in the probability of having an STI test by 6-weeks for e-STI versus walk-in STI testing among their clinic patients? This parameter can be written as
\[\psi = E[Y^1 | S=1] - E[Y^0 | S=1]\]
with the observed clinic and trial data consisting of $(V_i, W_i, S_i=1)$ and $(Y_i, A_i, V_i, W_i, S_i=0)$, respectively.

To mimic the described scenario, we use data from the \textit{GetTested} trial ($n_0=2063$) \cite{wilson_internet-accessed_2017}, a randomized trial comparing e-STI versus walk-in STI testing text messages. The outcome of the trial was any STI test within 6 weeks of randomization. To focus our illustration, observations were treated as marginally randomized (\textit{GetTested} used stratified randomization with different allocation probabilities, so analyses must incorporate this to be unbiased) and observations with missing outcome data (324, 16\%) were ignored. This data is referred to as the `full data'. For simplicity, only the assigned trial arm, outcome, age, and gender were considered. To induce non-positivity, we restricted the \textit{GetTested} trial data to only men (`restricted data', $n_0=1016$, 49\%). Because the \textit{GetTested} trial included women, we can benchmark the different approaches. As a stand-in for the clinic population, we simulated a data set ($n_1=1000$) where age and gender distributions differed from the \textit{GetTested} trial (Table \ref{tab1}).

\begin{table}[h]
	\caption{Descriptive statistics for the clinic and restricted trial data}
	\centering
	\begin{tabular}{lcccc}
		\hline
		& \multicolumn{2}{c}{Trial ($n_2=1016$)\textsuperscript{*}} &  & Clinic ($n_1=1000$) \\
		& e-STI               & Walk-in           &  &                     \\ \cline{2-3} \cline{5-5} 
		Age\textsuperscript{†}                     & 23 {[}20, 26{]}     & 22 {[}22,26{]}    &  & 21 {[}19,24{]}      \\
		Female                   & 0 (0\%)             & 0 (0\%)           &  & 574 (57\%)          \\
		Any STI test at 6 weeks\textsuperscript{‡} & 289 (53\%)          & 99 (21\%)         &  & -                   \\ \hline
	\end{tabular}
	\floatfoot{STI: sexually transmitted infection. \\
	* The \textit{GetTested} trial restricted to only males.\\
	† The first number is the median, with the bracketed numbers being the 25th and 75th percentiles, respectively.\\
	‡ Any STI test completed at 6 weeks post randomization.}
	\label{tab1}
\end{table}

\subsection*{Nonparametric Point Identification}

To proceed, $\psi$ is written in terms of the observable data without placing parametric constraints on the relationships between variables, referred to as \textit{nonparametric point identification}. For the randomized trial, we rely on the following assumptions
\[Y_i = Y_i^a \text{ if } a=A_i\]
\[E[Y^a | S=0] = E[Y^a | A=a, S=0] \text{ for } a \in\{0, 1\}\]
\[\Pr(A=a | S=0) > 0 \text{ for } a\in\{0, 1\}\]
which correspond to causal consistency, marginal treatment exchangeability, and treatment positivity, respectively \cite{zivich_positivity_2022, hernan_estimating_2006, cole_consistency_2009}. These assumptions are given by design for a marginally randomized trials. To transport the trial results to the clinic population \cite{westreich_transportability_2017}, the following assumptions are considered
\begin{equation}
	\begin{aligned}
		E[Y^a | S=1,V=v,W=w] = E[Y^a | S=0,V=v,W=w] \\
		\text{ for all } v,w \text{ where } \Pr(V=v,W=w | S=1) > 0
	\end{aligned}
	\label{exch1}
\end{equation}
\begin{equation}
	\Pr(V=v, W=w | S=0) > 0 \text{ for all } v,w \text{ where } \Pr(V=v,W=w | S=1) > 0
	\label{pos1}
\end{equation}
where \ref{exch1} is sample exchangeability (i.e., conditional exchangeability between the trial and target populations) and \ref{pos1} is sample positivity. As the target population consists of both men and women but the trial was limited to only men, assumtion \ref{pos1} is violated. In words, sample positivity holds by age but not by gender. As such, $\psi$ is \textit{not} nonparametrically point identified.

\section*{Potential Solutions}

While one always has the option to state that the question cannot be satisfactorily addressed given the data, we instead consider three solutions to address the motivating research question: restricting the target population to only men, restricting the covariate set for exchangeability between the trial and clinic such that we no longer need to account for gender, and synthesizing statistical and simulation models. To aid with intuition, the parameter of interest is re-expressed using the law of total expectation
\begin{equation}
	E[Y^a | S=1] = E[Y^a | S=1,W=0]\Pr(W=0 | S=1) +  E[Y^a | S=1,W=1]\Pr(W=1 | S=1)
	\label{parameter}
\end{equation}
As $\Pr(W=w | S=1)$ is identified given the clinic data and $E[Y^a | S=1, W=0]$ is identified given the trial data, the challenge is how to identify $E[Y^a | S=1, W=1]$.

\subsection*{Solution 1: Restrict the Target Population}

As $E[Y^a | S=1, W=1]$ is not identifiable from the trial, we could revise the parameter of interest to be the average causal effect among \textit{only men} at the clinic, $\psi' = E[Y^1 - Y^0 | S=1,W=0]$. Effectively, the original research question is modified by restricting the target population to regions where positivity is met. Identification of this revised parameter replaces \ref{exch1} and \ref{pos1} with
\begin{equation}
	\begin{aligned}
		E[Y^a | S=1,V=v,W=0] = E[Y^a | S=0,V=v,W=0]  \\
		\text{ for all } v \text{ where } \Pr(V=v | S=1,W=0) > 0
	\end{aligned}
	\label{exch2}
\end{equation}
\begin{equation}
	\Pr(V=v | S=0, W=0) > 0 \text{ for all } v \text{ where } \Pr(V=v | S=1,W=0) > 0
	\label{pos2}
\end{equation}
Here, \ref{pos2} is not violated, so $\psi'$ is nonparametrically point identified. For consistent estimation, either g-computation or inverse probability weighting (IPW) estimators can be applied (see Appendix A) \cite{westreich_transportability_2017, dahabreh_extending_2020}.

While $\psi'$ is identified, it is distinct from $\psi$. Specifically, this revised analysis no longer addresses the original research question as only a portion of the original target population is considered. If one (intentionally or mistakenly) assumes $\psi = \psi'$, then one implicitly assumes that men can directly stand-in for women and age distributions between populations do not differ by gender. Regardless, decisions regarding STI testing at the clinic must be made for women (i.e., avoidance of a decision is still a decision). The unwillingness to address the original question may lead the clinic to restrict e-STI testing offers to men. If e-STI testing is as or more effective in women than men, our analytical decision would then lead to a potential increase or perpetuation of inequalities in STI testing uptake by gender. Alternatively, if e-STI dissuades women but not men from completing STI tests, then adoption e-STI testing for all clinic patients could similarly result in inequalities. By restricting the target population, this analysis provides limited guidance  for the clinic.

\subsection*{Solution 2: Restricted Covariates for Exchangeability}

A second option is to modify the conditional exchangeability assumption in \ref{exch1} by replacing the adjustment set $\{V,W\}$ with $\{V\}$. Therefore, \ref{exch1} and \ref{pos1} are replaced by
\begin{equation}
	E[Y^a | S=1,V=v] = E[Y^a | S=0,V=v] \text{ for all } v \text{ where } \Pr(V=v | S=1) > 0
	\label{exch3}
\end{equation}
\begin{equation}
	\Pr(V=v | S=0) > 0 \text{ for all } v \text{ where } \Pr(V=v | S=1) > 0
	\label{pos3}
\end{equation}
This revised exchangeability assumption implies that the effect of STI testing options is homogenous on the difference scale by gender \cite{webster-clark_directed_2021}. Therefore, $\psi$ for the target population is identified under these revised assumptions. Again, either g-computation or IPW estimators could be used \cite{westreich_transportability_2017, dahabreh_extending_2020}.

Restricting the covariate set to age seems unjustified, especially if background knowledge led us to include gender in the set of covariates for conditional exchangeability between the trial and clinic originally. Further, this analysis has men directly `stand-in' for women, which leads to similar concerns to those discussed in solution 1 when effectiveness differs by gender. However, solution 2 does allow for age distributions to differ by gender, unlike solution 1 when $\psi = \psi'$ is assumed.

\subsection*{Solution 3: Model Synthesis}

As a third option, we propose a synthesis of statistical and simulation (e.g., mathematical, mechanistic \cite{lessler_mechanistic_2016, kirkeby_practical_2021}, agent-based \cite{railsback_agent-based_2011, el-sayed_social_2012}, microsimulation \cite{krijkamp_microsimulation_2018, caglayan_microsimulation_2018, grummon_health_2019}) models, where the simulation model is denoted by $r(a,W_i=1;\beta)$. Importantly, this simulation model is based on knowledge external from the available data. Here, $\beta$ is assumed to be unknown but a distribution of plausible values are considered feasible to specify (denoted as $\tilde{\beta}$). To maintain the generality of our discussion, how a simulation model and $\tilde{\beta}$ are specified are left till the end of the section.

\subsubsection*{G-computation}

First, we propose an outcome modeling approach akin to standard g-computation \cite{snowden_implementation_2011}. To generate predicted values of $Y$ under $a$, a combination of statistical and simulation modeling is used
\[f_a(V_i,W_i;\alpha,\beta) = t\{q(a,V_i,W_i=0;\alpha) + r(a, W_i=1;\beta)\}\]
where $q(a,V_i,W_i=0;\alpha)$ is a statistical model for $E[Y | A_i = a, V_i, W_i=0, S_i=0]$ and $t()$ is a generic transformation (e.g., identity for linear models, inverse logit for logistic models). Here, the simulation model, $r(a, W_i=1;\beta)$, shifts the fitted statistical model to generate predicted values of $Y^a$ for women. The synthesis g-computation estimator for $\psi$ is 
\[\hat{\psi}_g = \frac{1}{n_1} \sum_{i=1}^{n} \{f_1(V_i,W_i;\hat{\alpha}, \tilde{\beta}) I(S_i = 1)\} - \frac{1}{n_1} \sum_{i=1}^{n} \{f_0(V_i,W_i;\hat{\alpha}, \tilde{\beta}) I(S_i = 1)\}\]
where $n$ is the number of observations in the combined clinic and trial data, and $n_1 = \sum_{i=1}^{n} I(S_i = 1)$. To apply this estimator, first the nuisance parameters for the statistical model, $\alpha$, are estimated using the trial data. Keeping with the motivating example, a logistic regression model could be fit predicting STI test completion among men. Next, a simulation model is developed to update the fitted statistical model values for women. Then $\hat{\psi}_g$ is computed based on $f_a(V_i,W_i;\hat{\alpha}, \tilde{\beta})$. To incorporate the uncertainty of $\hat{\alpha}$, $\tilde{\beta}$, and sampling of the target population, we propose the following semiparametric bootstrap or Monte Carlo procedure. First, randomly draw $\tilde{\beta}^*$ from the specified distribution for $\tilde{\beta}$, and $\hat{\alpha}^*$ from a multivariate normal distribution based on $\hat{\alpha}$ and the estimated covariance matrix for $\hat{\alpha}$. Next, resample with replacement the clinic data. Using $\hat{\alpha}^*$, $\tilde{\beta}^*$, and the resampled clinic data, compute $\hat{\psi}_g$. This process is then repeated many times (at least $10^5$). This approach to uncertainty analysis has also been used in probabilistic bias analyses \cite{fox_presentation_2021}. Results from all the repetitions can then be numerically summarized (e.g., mean and standard deviation, percentiles) or presented visually (e.g., histogram, violin plot).

In the illustrative example, we could consider the following model
\begin{equation*}
	\begin{aligned}
		f_a(V_i, W_i; \alpha, \beta) = & \text{expit}\{q(a,V_i, W_i=0;\alpha) + r(a,W_i=1;\beta)\} \\
		= & \text{expit}\{\alpha_0 + \alpha_1 a + \alpha_2 V_i + \beta_0 W_i + \beta_1 a W_i\}
	\end{aligned}
\end{equation*}
Note that this model assumes no three-way interactions (i.e., $\beta_2 a W_i V_i$ is excluded), which may be reasonable to assume in applications \cite{seamans_general_2021}. One challenge to application of this estimator is that the direction of conditional coefficients, like $\beta$, can be counter-intuitive, as shown by the Table 2 Fallacy \cite{westreich_table_2013, bandoli_revisiting_2018, williamson_factors_2020, westreich_comment_2021, keller_rates_2018}. This concern is compounded by the best available external information for simulation models often being marginal parameters \cite{murray_comparison_2017, murray_challenges_2020}. Therefore, specifying $\beta$ for the synthesis g-computation estimator may be difficult.

\subsubsection*{Inverse Probability Weighting}
To avoid conditional parameters in simulation models, we propose a synthesis IPW estimator,
\[\hat{\psi}_w = \frac{1}{n_1} \sum_{i=1}^{n} 
\{h_1(V_i,W_i;\hat{\gamma}, \tilde{\delta}, \hat{\eta}) I(S_i = 1)\} - 
\frac{1}{n_1} \sum_{i=1}^{n} \{h_0(V_i,W_i;\hat{\gamma}, \tilde{\delta}, \hat{\eta}) I(S_i = 1)\}\]
where
\[h_a(V_i,W_i;\gamma, \delta, \eta) = t\{q(a,V_i,W_i=0;\gamma, \eta) + r(a,W_i=1;\delta)\}\]
Here, the synthesis model consists of a marginal structural model. For the illustrative example, the following logistic marginal structural model for men could be specified
\[t\{q(a,V_i,W_i=0;\gamma, \eta)\} = \text{expit}\{\gamma_0 + \gamma_1 A_i\}\]
where $\eta$ are nuisance parameters for the inverse probability weights in the statistical model. Therefore, the synthesis IPW model would be
\begin{equation*}
	\begin{aligned}
		h_a(V_i,W_i;\gamma, \delta, \eta) = & \text{expit}\{q(a,V_i,W_i=0;\gamma, \eta) + r(a,W_i=1;\delta)\} \\
		= & \text{expit}\{\gamma_0 + \gamma_1 a + \delta_0 W_i + \delta_1 a W_i\}
	\end{aligned}
\end{equation*}
As shown, the simulation model parameters, $\delta$, are not conditional on $V_i$.

To implement this estimator, the overall inverse probability weights are estimated. The weights for the illustrative example are
\[\frac{1}{\Pr(A_i=a | S_i = 0)} \times \frac{\Pr(S_i = 1 | V_i, W_i = 0)}{\Pr(S_i = 0 | V_i, W_i = 0)}\]
where a logistic model can be used to estimate the probability of membership in the clinic data restricted to men, $\Pr(S_i = 1 | V_i, W_i = 0)$ \cite{westreich_transportability_2017}. The parameters of a marginal structural model can then be estimated using weighted maximum likelihood estimation with the clinic data. For point and variance estimation, an analogous Monte Carlo procedure with $\hat{\gamma}$ and $\tilde{\delta}$ can be used.

While the synthesis IPW estimator avoids the simulation parameters being conditional on $V$, $\delta$ are assumed to originate from a population with a similar age distribution to the clinic. If the distribution for $\delta$ was based on a population with a different age distribution, then biased estimates of $\psi$ could result. While differences in the age distribution could be accounted for through a more richly specified simulation model (i.e., including $\delta_2 a W_i V_i$), selection of the simulation model parameters for such models may be difficult.

\subsubsection*{Building a Simulation Model}

A primary challenge for application of the proposed estimators is the development of a simulation model. As the simulation model and corresponding parameters are inestimable given the available data, external information must be relied upon. While building a simulation model may appear to entail stronger assumptions relative to restricting the target population or set of covariates, complexity of implementation is often confused with strength of assumptions. Here, both restriction approaches are special cases of the synthesis approach. Specfically, both assume a simulation model that always contributes zero to $f_a$ or $h_a$. When viewed in this manner, approaches based on restricting the population or covariate set entail more constraining assumptions. So, even if external knowledge is limited or weak, model synthesis offers a way to address the original question under more flexible assumptions than standard approaches.

To guide our discussion of building a simulation model, we consider how one may go about this task for the motivating example. The simulation model needs to generate the probability of STI test uptake by offered testing option among women. This can be accomplished in various ways. As done above, the probability of STI test uptake among women could be directly simulated by adjusting a logistic regression model fit to men. Alternatively, the probability of uptake among women could be simulated through a series of intermediate mechanisms (e.g., probability of reading text, probability of completing online test order versus traveling to clinic). Deciding between modeling strategies should be based on the types of external information available. For example, if external information existed comparing STI testing options between men and women in other contexts then the first simulation modeling option might be preferred. To obtain information to construct a simulation model, a diverse set of external sources should be considered (e.g., trial, observational studies, studies on other treatments with similar mechanisms of action, pharmacokinetic studies, animal models). Additionally, a panel of subject-matter experts could be assembled to summarize the current evidence base \cite{bojke_good_2021, ohagan_expert_2019}. Expert knowledge has been used in a similar way to develop sensitivity analyses \cite{shepherd_does_2008}. Importantly, simulation model parameters should reflect both the uncertainty in the chosen model and variability of the information sources. More extensive discussions on building simulation models exist as part of a rich literature in both epidemiology and disease ecology \cite{krijkamp_microsimulation_2018, roberts_conceptualizing_2012, railsback_agent-based_2011, slayton_modeling_2020}. For building complex models in practice, seeking out modeling-specialist collaborators is highly advised.

\section*{Application}

As stated previously, the restricted trial and clinic data (Table \ref{tab1}) are used to estimate the difference in probability of 6-week STI test uptake between text messages offering e-STI versus walk-in STI testing in the clinic population. Analyses used Python 3.9.5 (Beaverton, OR) with the following packages: \texttt{NumPy} \cite{harris_array_2020}, \texttt{SciPy} \cite{virtanen_scipy_2020}, \texttt{pandas} \cite{mckinney_data_2010}, and \texttt{delicatessen} \cite{zivich_delicatessen_2022}. The \textit{GetTested} data is available from Wilson et al. \cite{wilson_internet-accessed_2017}, and code is provided at github.com/pzivich/publications-code.

Age was included in all nuisance models with both a linear and quadratic term, and no interaction terms. As a benchmark, we present results that used the full data. The remainder of analyses used the restricted data. Details on the standard g-computation and IPW estimators are provided in Appendix A. For the restricted target population approach, estimators were applied to the clinic data include both men and women. For point and variance estimation, we used M-estimation \cite{zivich_delicatessen_2022, stefanski_calculus_2002}.

For the synthesis approach, the described g-computation and IPW estimators were applied with 10,000 Monte Carlo iterations. The synthesis models were 
\[f_{A_i}(V_i, W_i; \alpha, \beta) = \text{expit}(\alpha_0 + \alpha_1 A_i + \alpha_2 V_i + \alpha_3 V_i^2 + \beta_0 W_i + \beta_1 A_i W_i)\]
for g-computation and 
\[h_{A_i}(V_i, W_i; \alpha, \beta) = \text{expit}(\gamma_0 + \gamma_1 A_i + \delta_0 W_i + \delta_1 A_i W_i)\]
for the IPW estimator. Different specifications for $\beta$ and $\delta$ were compared (Table \ref{tab2}). The first assumed a strict null (i.e., no effect with certainty), where results from the restricted covariate set and model synthesis were expected to match (up to Monte Carlo error). The second set proposed an uncertain null with a trapezoidal distribution ranging from -2 to 2 and was uniform between -1 to 1 \cite{fox_method_2005}. The third set corresponded to a case where accurate (valid and precise) outside information was available. As a stand-in, we estimated the coefficients of the synthesis model above using the full \textit{GetTested} data. The corresponding $\beta$ or $\delta$ estimates were then used as inputs for the simulation model. This choice of simulation model parameters is expected to provide a similar point estimate to the full data, but variance estimates can differ due to the covariance between simulation model parameters being ignored and additional uncertainty in the estimated statistical model parameters in the restricted data. Next, the $\beta$ and $\delta$ inputs from the previous scenario were reversed to demonstrate the issue of confusing marginal and conditional parameters. Finally, the fifth set multiplied the previous by negative one to correspond to inaccurate external information.

\begin{table}[]
	\caption{Simulation model parameter distributions}
	\begin{tabular}{llcc}
		\hline
		&                        & $\beta_0$ or $\delta_0$                     & $\beta_1$ or $\delta_1$                     \\ \cline{3-4} 
		\multicolumn{2}{l}{G-computation} &                                             &                                             \\
		& Strict null            & 0                                           & 0                                           \\
		& Uncertain null         & $\text{Trapezoid}(-2,-1,1,2)$               & $\text{Trapezoid}(-2,-1,1,2)$               \\
		& Accurate\textsuperscript{*}              & $\text{Normal}(\mu=-0.0160, \sigma=0.1761)$ & $\text{Normal}(\mu=-0.6270, \sigma=0.2227)$ \\
		& Inaccurate\textsuperscript{†}             & $\text{Normal}(\mu=0.0160, \sigma=0.1761)$  & $\text{Normal}(\mu=0.6270, \sigma=0.2227)$  \\
		\multicolumn{2}{l}{IPW}           &                                             &                                             \\
		& Strict null            & 0                                           & 0                                           \\
		& Uncertain null         & $\text{Trapezoid}(-2,-1,1,2)$               & $\text{Trapezoid}(-2,-1,1,2)$               \\
		& Accurate\textsuperscript{*}              & $\text{Normal}(\mu=0.1380, \sigma=0.1931)$  & $\text{Normal}(\mu=-0.6914, \sigma=0.2460)$ \\
		& Inaccurate\textsuperscript{†}             & $\text{Normal}(\mu=-0.1380, \sigma=0.1931)$ & $\text{Normal}(\mu=0.6914, \sigma=0.2460)$  \\ \hline
	\end{tabular}
	\floatfoot{IPW: inverse probability weighting. \\
	* The accurate simulation model inputs were based on fitting corresponding statistical models using the full \textit{GetTested} data set.\\
	† The inaccurate simulation model inputs were based on the accurate parameters but flipping the sign for the center of the distribution.}
	\label{tab2}
\end{table}

\subsection*{Results}

\begin{figure}
	\centering
	\caption {G-computation results for transportation of the average causal effect to the clinic.}
	\includegraphics[width=0.9\linewidth]{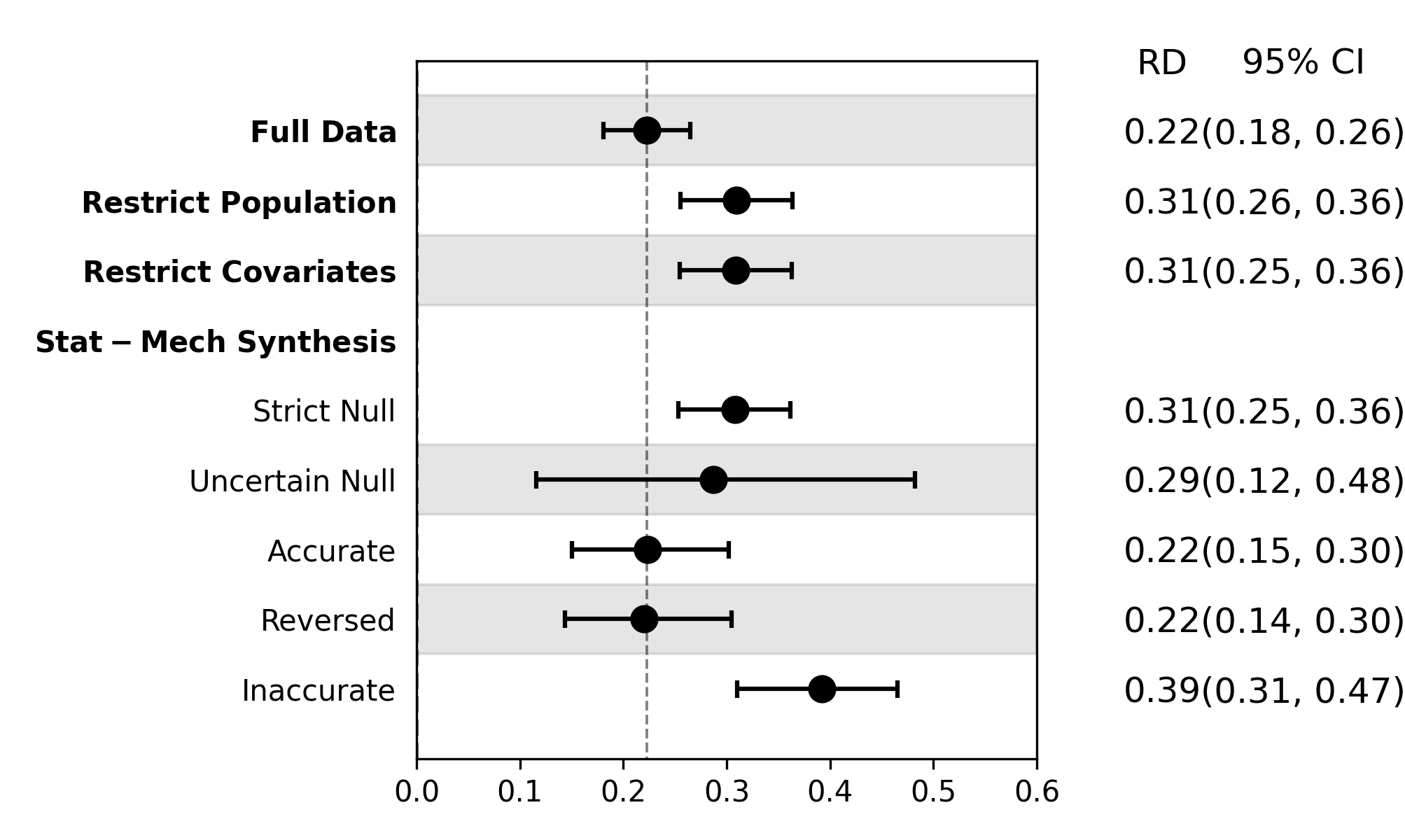}
	\floatfoot{RD: risk difference, 95\% CI: confidence interval. Dashed vertical line indicates the full data point estimate.}
	\label{fig1}
\end{figure}

\begin{figure}
	\centering
	\caption {Inverse probability weighting estimator results for transportation of the average causal effect to the clinic.}
	\includegraphics[width=0.9\linewidth]{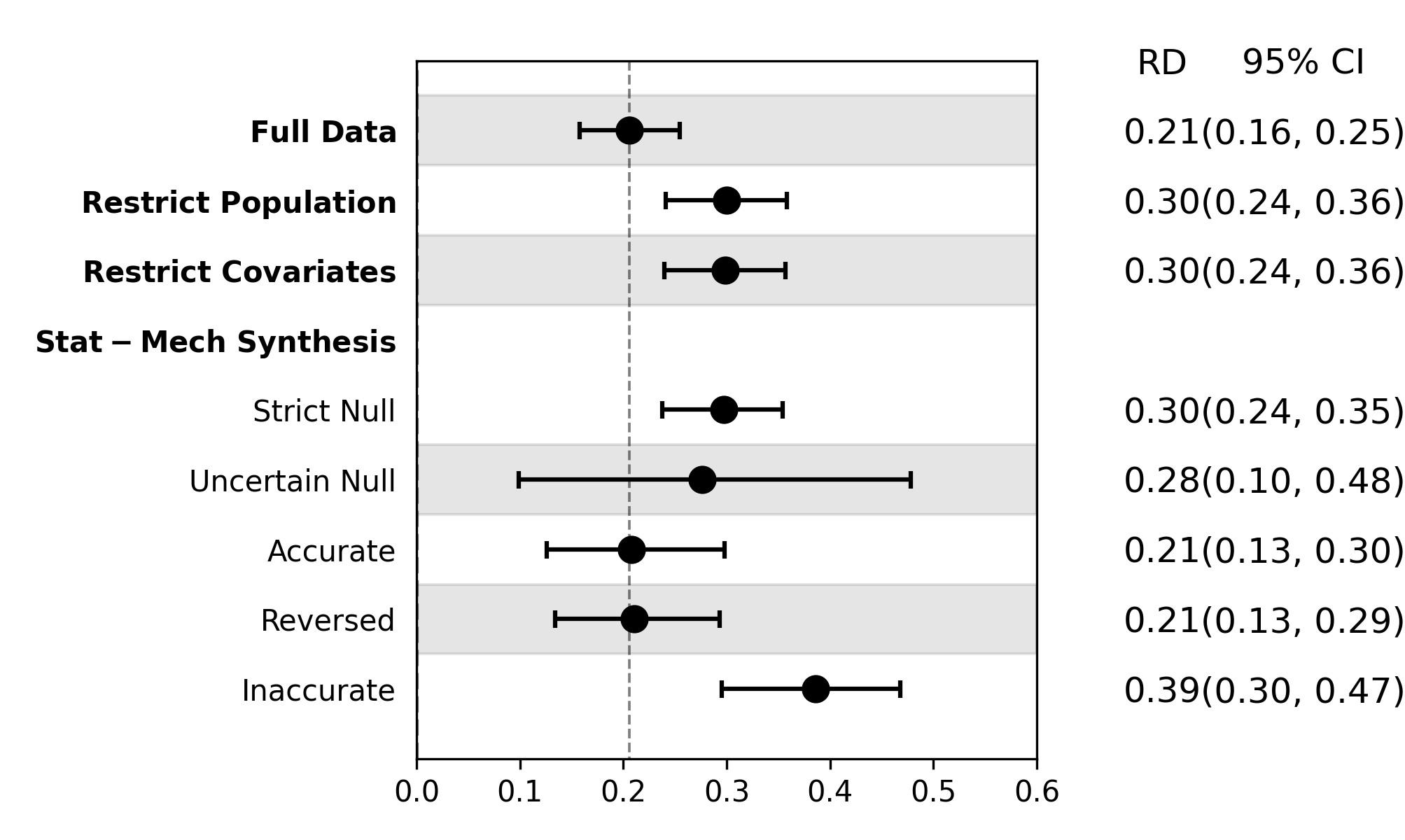}
	\floatfoot{RD: risk difference, 95\% CI: confidence interval. Dashed vertical line indicates the full data point estimate.}
	\label{fig2}
\end{figure}

Results are presented in Figures \ref{fig1} and \ref{fig2}, respectively. Here, the restricted target population and restricted covariate set approaches similarly overestimated the effectiveness of offering e-STI testing for the clinic population, as offering e-STI testing was more effective among men. The similar results were attributable to men and women having the same age distribution in the clinic population. For the model synthesis, results were dependent on the choice of simulation model parameters, as expected. When the simulation model assumed a strict null, the results were nearly identical to the restriction approaches. With uncertain null distributions, results were less precise but e-STI testing remained beneficial. In the setting with accurate external knowledge, the synthesis approach gave a nearly identical point estimate to the full data approach, but with reduced precision. When the specified marginal and conditional parameters were reversed, results were similar to the accurate specification, likely due to marginal and conditional parameters being similar in this example (Table \ref{tab2}). In the setting with inaccurate knowledge, the synthesis approach over-estimated the effectiveness of e-STI testing.

\subsection*{Simulations}

To further explore the operating characteristics of the different approaches, a simulation study based on the illustrative example was conducted (detailed in Appendix B). As expected, both restriction approaches were biased and had poor confidence interval coverage. Both g-computation and IPW synthesis estimators performed well unless simulation model parameters were grossly misspecified. Relations between the difference approaches also aligned with theoretical expectations. Importantly, the synthesis approach had wider confidence intervals than both restriction approaches when simulation model parameters were drawn from distributions, reflecting uncertainty in simulation model parameters.

\section*{Discussion}

Here, we examined three solutions to dealing with positivity violations by a binary covariate in transportability problems: restrict the target population, restrict the covariate set for exchangeability, and a synthesize of statistical and simulation models. Synthesis models allow epidemiologists to address questions where positivity is violated (unlike restricting the target population) but avoid modifying the conditional exchangeability assumption (unlike restricting the covariate set). Even if one is willing to assume null or homogeneous effects across non-positive regions, model synthesis can incorporate uncertainty into this assumption.

Here, violations of the positivity assumption for transportability by an observed covariate were considered. Other work on relaxing transportability and generalizability identification assumptions has instead considered violations of the exchangeability assumption by unmeasured covariates \cite{nilsson_proxy_2023, nguyen_sensitivity_2017, lesko_effect_2016}. For some causal structures, a pair of proxy variables can be leveraged for identification despite an unmeasured outcome predictor that differs between populations \cite{nilsson_proxy_2023}. However, a positivity assumption is needed for the unmeasured predictor and proxy variables, in addition to other assumptions \cite{tchetgen_introduction_2020, zivich_introducing_2023}. Other work has taken a sensitivity analysis approach, where unobserved covariates are predicted from a statistical model fit to the non-target population sample and then shifted by data-dependent values \cite{nguyen_sensitivity_2017, lesko_effect_2016}. These latter approaches share similarities to the implemented synthesis estimators, where statistical model predictions are shifted by a simulation model. This overlap suggests that synthesis approaches could be used to jointly address exchangeability and positivity violations.

Previous epidemiologic research on statistical and simulation modeling has largely focused on comparing them, or triangulation of their results \cite{murray_comparison_2017, mooney_g-computation_2021, murray_emulating_2021, ip_reconciling_2013, hernan_invited_2015, buchanan_disseminated_2021, el-sayed_social_2012, lofgren_re_2017, ackley_dynamical_2022, halloran_simulations_2017, keyes_invited_2017, edwards_invited_2017, lofgren_mathematical_2014, arnold_dag_2019, ackley_compartmental_2017, mooney_g-computation_2022}; but less epidemiologic research has explicitly described how these methods can be integrated together \cite{cerda_systems_2019}. Regardless, a variety of approaches can be framed as syntheses of statistical and simulation modeling, with notable examples including quantitative bias analysis, sensitivity analysis, Bayesian methods, and measurement error correction \cite{greenland_bayesian_2009, greenland_interval_2004, robins_sensitivity_2000, cole_sensitivity_2023}. Outside of epidemiology, research in physics, engineering, and climate science have used hybrids of statistical and simulation modeling (often referred to as `phenomenological' and `mathematical' models, respectively) \cite{rahmstorf_semi-empirical_2007, wright_semi-empirical_nodate, sausen_efficiency_2018, rezaei_hybrid_2020}.

Our discussion focused on point identification, but one could instead opt for partial identification via nonparametric bounds \cite{manski_nonparametric_1990, cole_nonparametric_2019}, which avoids exchangeability and positivity assumptions. Here, we could have calculated the lower and upper bounds for pieces of $\psi$ that violated the positivity assumption by predicting either 0 or 1 for all women, respectively. These can then be combined to get lower and upper bounds for $\psi$. Importantly, the proposed synthesis g-computation and IPW estimators correspond to these bounds when the simulation model always generates predictions of 0 (or 1) for women. Therefore, the synthesis approach can also be viewed as a way to narrow the nonparametric bounds via external information.

Future areas ripe for extending the proposed synthesis framework include addressing other sources of systematic errors and non-positivity by continuous covariates, alternative estimators, and more precisely describing the inferential model. Here, we focused on transportability but conditional exchangeability and positivity assumptions are commonly used to address other systematic errors, including confounding, missing data, and measurement error \cite{zivich_positivity_2022}. Returning to the example, observations in the \textit{GetTested} trial with missing outcomes were dropped, but this is not recommended in practice \cite{ware_missing_2012}. To account for missing outcomes by observed covariates, one could consider multiple imputation or inverse probability of missingness weights \cite{perkins_principled_2018}. If missingness depended on unobserved covariates or the missing values themselves; then sensitivity analyses \cite{robins_sensitivity_2000, greenland_basic_1996, cole_sensitivity_2023}, or extensions of the proposed synthesis could be applied. Similarly, one can consider extending synthesis models for non-positivity by continuous covariates. While some extensions may be trivial, there are important details to consider. For example, conditional exchangeability for average causal effect requires as symmetric positivity assumption, not present in sample positivity \cite{zivich_positivity_2022, zivich_use_2022}. Additionally, positivity for some systematic errors may remain uncorrectable (e.g., the average causal effect of pregnancy on blood pressure in a population that includes persons unable to become pregnant). Second, alternative estimators might be of interest. Variations on the statistical models, simulation models, or Monte Carlo procedure are possible, with particular interest in machine learning and multiply robust estimation \cite{bang_doubly_2005, zivich_machine_2021}. Further, testable implications for simulation models, similar to the natural course for the g-formula \cite{keil_parametric_2014} or balance with IPW \cite{austin_moving_2015}, are of interest. Finally, inference for model synthesis was left largely unspecified. While we refer to the intervals as `confidence intervals', the model synthesis intervals do not have the same operational characteristics as standard frequentist confidence intervals since both the random error of the statistical model and uncertainty in simulation model is reflected. Instead, we suggest intervals are interpreted in terms of their precision or width \cite{poole_low_2001}, an approach used for interval interpretation generally (including those lacking frequentist interpretations \cite{greenland_interval_2004, greenland_bayesian_2006}). Alternatively, one could adopt a semi-Bayes interpretation, as done in probabilistic bias analysis \cite{fox_presentation_2021}. For a fully Bayesian interpretation, several modifications to the synthesis approach would be needed. First, $\psi$ would instead represent a random variable, which has consequences for identification \cite{li_bayesian_2023}. Second, as explained elsewhere \cite{robins_bayesian_2015}, the IPW estimator is not amenable to a fully Bayesian framework. Finally, estimation requires specification of priors for the statistical model parameters.

Decisions must be made, with or without complete data. The aspiration of the data alone guiding all actions is an impossible dream \cite{quine_main_1951, robins_impossibility_1999, fajardo-fontiveros_fundamental_2022}. To make any reasonable progress in learning about causal effects, we depend on relevant knowledge external to the data \cite{robins_data_2001}. Within transportability and causal inference more generally, external information has often been limited to the relations between variables (i.e., arrows in a causal diagram) and functional forms in statistical models. Here, we extended this idea to include external knowledge to address positivity violations.

\section*{Acknowledgments}

This work was supported in part by T32-AI007001 (PNZ), R01-AI157758 (JKE, BES, SRC), R01-GM140564 (JKE, JL), and R35-GM147013 (ETL).
Corresponding code is available at https://github.com/pzivich/publications-code

\small
\bibliographystyle{ieeetr}
\bibliography{biblio}{}

\newpage 

\normalsize

\section*{Appendix}

\subsection*{Appendix A: Estimators}

In the applied example and simulations, M-estimation is used for the restricted population and restricted covariate set estimators. Briefly, an M-estimator, denoted as $\hat{\theta}$, is the solution for $\theta$ in the estimating equation, $\sum_{i=1}^{n} \varphi(O_i;\theta) = 0$; where $O_i$ is the observed data for $n$ independent and identically distributed units, $\theta$ is a $k$ dimensional vector of parameters, and $\varphi$ is a known $k$-dimensional vector of estimating functions that do not depend on $i$ \cite{zivich_delicatessen_2022, stefanski_calculus_2002}. The variances of the parameter estimates can be estimated using the empirical sandwich variance estimator
\[\mathbb{V}(O_i; \hat{\theta}) = \frac{1}{n} \left[\mathbb{B}(O_i; \hat{\theta})^{-1} \mathbb{F}(O_i; \hat{\theta}) \{\mathbb{B}(O_i; \hat{\theta})^{-1}\}^{T}\right]\]
where the `bread' is
\[\mathbb{B}(O_i; \hat{\theta}) = \frac{1}{n} \sum_{i=1}^{n} \{- \varphi(O_i; \hat{\theta})\}\]
with $\varphi'(O_i;\hat{\theta})$ indicating the matrix of partial derivatives (i.e., Jacobian), and the `filling' is
\[\mathbb{F}(O_i; \hat{\theta}) = \frac{1}{n} \sum_{i=1}^{n} \{\varphi(O_i; \hat{\theta}) \varphi(O_i; \hat{\theta})^T\}\]
Importantly, the sandwich estimator allows for the propagation of uncertainty between parameters that depend on each other. In short, a consistent variance can be easily estimated while avoiding more computationally demanding procedures, like the nonparametric bootstrap.

\subsubsection*{Restrict the Target Population}

For the restricted target population g-computation estimator, the stacked estimating functions were
\[\varphi(O_i; \theta) = 
\begin{bmatrix}
	I(S_i = 0) \left(\left[ Y_i - \text{expit}\left\{g(A_i,V_i)^T \alpha \right\} \right] g(A_i, V_i)\right)\\
	I(S_i = 1, W_i = 0) \left[\text{expit}\left\{g(0,V_i)^T \alpha\right\} - \theta_0\right] \\
	I(S_i = 1, W_i = 0) \left[\text{expit}\left\{g(1,V_i)^T \alpha\right\} - \theta_1\right] \\
	(\theta_1 - \theta_0) - \theta_2
\end{bmatrix}\]
where $\theta = (\alpha, \theta_0, \theta_1, \theta_2)$, $g(A_i,V_i) = (1, A_i, V_i, V_i^2)$ is the design matrix and $\text{expit}(b) = \frac{\exp(b)}{1 + \exp(b)}$. The first estimating function is a logistic regression model for $Y_i$ among the trial data. The second and third estimating functions are the predicted mean under $A_i=0$ and $A_i=1$ among men in the clinic data, respectively. The last estimating function is for the risk difference.

The inverse probability weighting (IPW) estimator consisted of the following stacked estimating functions
\[\varphi(O_i; \theta) = 
\begin{bmatrix}
	I(S_i=0) \left\{A_i - \text{expit}(\mu)\right\} \\
	I(W_i = 0) \left(\left[S_i - \text{expit}\left\{l(V_i)^T \sigma\right\}\right] l(V_i) \right) \\
	(Y_i - \theta_0) \frac{I(S_i = 0) \text{expit}\left\{l(V_i)^T \sigma \right\} }{1 - \text{expit}\left\{l(V_i)^T \sigma \right\}} \frac{I(A_i = 0)}{1 - \text{expit}(\mu)} \\
	(Y_i - \theta_1) \frac{I(S_i = 0) \text{expit}\left\{l(V_i)^T \sigma \right\} }{1 - \text{expit}\left\{l(V_i)^T \sigma \right\}} \frac{I(A_i = 1)}{\text{expit}(\mu)} \\
	(\theta_1 - \theta_0) - \theta_2
\end{bmatrix}\]
where $\theta = (\mu, \sigma, \theta_0, \theta_1, \theta_2)$ and $l(V_i) = (1, V_i, V_i^2)$ is the design matrix. The first estimating function is an intercept-only regression model for assigned STI testing in the trial data. The second estimating function is a logistic model for the conditional probability of being in the clinic, fit using only men. The third and fourth estimating functions are Hajek estimators for $A_i=0$ and $A_i=1$, respectively. The last estimating function is for the risk difference.

\subsubsection*{Restricting the Covariate Set}

The g-computation estimator for the restricted covariate set was implemented via the following stacked estimating functions
\[\varphi(O_i; \theta) = 
\begin{bmatrix}
	I(S_i = 0) \left(\left[ Y_i - \text{expit}\left\{g(A_i,V_i)^T \alpha \right\} \right] g(A_i, V_i)\right)\\
	I(S_i = 1) \left[\text{expit}\left\{g(0,V_i)^T \alpha\right\} - \theta_0\right] \\
	I(S_i = 1) \left[\text{expit}\left\{g(1,V_i)^T \alpha\right\} - \theta_1\right] \\
	(\theta_1 - \theta_0) - \theta_2
\end{bmatrix}\]
The key difference from the previous g-computation estimating functions is predictions are generated in the second and third equations for both men and women.

The stacked estimating functions for the IPW estimator 
\[\varphi(O_i; \theta) = 
\begin{bmatrix}
	I(S_i=0) \left\{A_i - \text{expit}(\mu)\right\} \\
	\left[S_i - \text{expit}\left\{l(V_i)^T \sigma\right\}\right] l(V_i) \\
	(Y_i - \theta_0) \frac{I(S_i = 0) \text{expit}\left\{l(V_i)^T \sigma \right\} }{1 - \text{expit}\left\{l(V_i)^T \sigma \right\}} \frac{I(A_i = 0)}{1 - \text{expit}(\mu)} \\
	(Y_i - \theta_1) \frac{I(S_i = 0) \text{expit}\left\{l(V_i)^T \sigma \right\} }{1 - \text{expit}\left\{l(V_i)^T \sigma \right\}} \frac{I(A_i = 1)}{\text{expit}(\mu)} \\
	(\theta_1 - \theta_0) - \theta_2
\end{bmatrix}\]
The key difference from the IPW estimator under the restricted target population was the second estimating function not being restricted to men (i.e., $W_i = 0$).

\subsection*{Appendix B: Simulations}

\subsubsection*{Data Generation}

Covariates for the clinic population ($S_i = 1$) were generated according to the following distributions
\[W_i \sim \text{Bernoulli}(0.667)\]
\[V_i \sim \text{Trapezoid}(\text{min}=18, \text{mode}_1 = 18, \text{mode}_2 = 25, \text{max} = 30)\]
$V_i$ was additional rounded to be an integer. Covariates for the trial population ($S_i=0$) were generated from
\[W_i \sim \text{Bernoulli}(0.0)\]
\[V_i \sim \text{Trapezoid}(\text{min}=18, \text{mode}_1 = 25, \text{mode}_2 = 30, \text{max} = 30)\]
with $V_i$ being rounded again. To summarize, the clinic population included men and women, and consisted of a younger population than the trial. Outcomes were generated from the following model
\[Y_i \sim \text{Bernoulli}(p_{Y_{i}})\]
\[p_{Y_i} = \text{expit}(-3.25 + 1.50 A_i + 0.08 - 0.02 W_i -0.65 A_i W_i)\]
The true value for $\psi$ (i.e., $0.216697$) was approximated by simulating the potential outcomes for 10 million observations with the clinic covariate pattern.

\subsubsection*{Methods}

The previously described estimating functions for the restricted g-computation and IPW estimators were applied in the simulations, with the following modifications to the design matrices for the corresponding nuisance models: $g(A_i, V_i) = (1, A_i, V_i)$ and $l(V_i) = (1, V_i, V_i I(V_i > 25))$. For model synthesis, the model for g-computation was
\[f_a(V_i, W_i; \alpha, \beta) = \text{expit}(\alpha_0 + \alpha_1 a + \alpha_2 V_i + \beta_0 W_i + \beta_1 A_i W_i)\]
and the model for the IPW estimator was
\[h_a(V_i, W_i; \gamma, \delta, \eta) = \text{expit}(\gamma_0 + \gamma_1 a + \delta_0 W_i + \delta_1 A_i W_i)\]
Point, lower, and upper confidence intervals were estimated by the median, 2.5\textsuperscript{th}, and 97.5\textsuperscript{th} percentiles, respectively, from 5000 repetitions of the semiparametric bootstrap. 

Seven sets of $\beta$ and $\delta$ specifications for each estimator were compared: strict null, uncertain null, accurate, inaccurate, accurate with covariance, reversed marginal and conditional parameters, and an external information source with a different age distribution. The strict null assumed all $\beta = \delta = 0$. Results for this specification are expected to match the restricted target population and covariate set. The uncertain null used a trapezoid distribution that is uniform from -1 to 1 and ranges from -2 to 2. 

For the accurate parameter specification, we used a `secret' trial as a stand-in for variable external knowledge. In tandem with the clinic and trial data sets, a secret second trial was conducted in each iteration of the simulation that included both men and women. By taking this approach, the simulation model parameters were unbiased in expectation but also vary within each iteration of the simulation study. Secret trials consisted of either 2000 or 4000 observations generated following the covariate pattern in the clinic population with $A_i$ randomly assigned and outcomes generated from the preceding model for $p_{Y_i}$. The secret trial was used to estimate $\hat{\beta}$, $\hat{\delta}$, $\widehat{Var}(\hat{\beta})$, and $\widehat{Var}(\hat{\delta})$; which were then provided to the model synthesis estimators. Importantly, no individual level data from the secret trial was used by the model synthesis estimators (i.e., synthesis estimators only saw the estimated parameters from the secret trial). The accurate specification is expected to be unbiased, but confidence interval coverage is expected to be conservative since the covariance between the parameters in $\hat{\beta}$ or $\hat{\delta}$ was ignored. The inaccurate parameter specification used the accurate specification parameters multiplied by negative one. This specification was expected to be biased due to the reversal of the direction of the relationships. The accurate with covariance parameters used the secret trial, but the estimated covariance matrices for $\hat{\beta}$ and $\hat{\delta}$ were used to draw parameters from a multivariate normal distribution (as opposed to independent normal distributions). As the covariance between parameters is not ignored, confidence intervals are expected to have nominal coverage. For the reversed specification, the marginal and conditional parameters were flipped with each other. Finally, the secret trial was conducted in a third population with the following age distribution, $\text{Trapezoid}(\text{min}=18, \text{mode}_1 = 29, \text{mode}_2 = 30, \text{max} = 30)$

Simulations consisted of 2000 iterations with $n_1 = 1000, n_0 = 1000$ and $n_1 = 1000, n_0 = 500$. Evaluation metrics were bias, confidence limit difference (CLD), and 95\% confidence interval coverage \cite{morris_using_2019}. As the variance estimates for the model synthesis approach were not necessarily normally distributed in all cases, we do not report the average standard error or the standard error ratio. Bias was defined as the mean of the difference between the estimate and true value. CLD was defined as the mean of the difference between the upper and lower confidence intervals, and is an indicator of precision (e.g., smaller CLD indicates greater precision). 95\% confidence interval coverage was defined as the proportion of intervals that contained the truth. Simulations were conducted using Python 3.9.5 (Beaverton, OR) with the following packages: \texttt{NumPy} \cite{harris_array_2020}, \texttt{SciPy} \cite{virtanen_scipy_2020}, \texttt{pandas} \cite{mckinney_data_2010}, and \texttt{delicatessen} \cite{zivich_delicatessen_2022}.

\subsubsection*{Results}

For $n_1 = 1000, n_0 = 1000$, the restricted target population and restricted covariate set were biased and had poor confidence interval coverage (Appendix Tables \ref{atab1} \& \ref{atab2}). Results were approximately equal across these approaches since the distribution of $V_i$ did not differ by $W_i$. As expected, performance of the model synthesis approach depended on the selection of parameters. 

For the strict null, results for g-computation and IPW were nearly identical to the restriction approaches. For the uncertain null, there was some bias, but 95\% confidence interval coverage was 100\%. This over-coverage results from the wide confidence interval, as indicated by the CLD. For the accurate specification, bias was near zero and confidence interval coverage was slightly above 95\% for both estimators. The accurate specification with the covariance matrix had near zero bias and 95\% confidence interval coverage, which is in line with the expected frequentist performance of the synthesis model with this simulation model specification. Results for both the reversed marginal and conditional parameters and different age distribution of the simulation model parameter source population had similar performance to the accurate specification. Overall, g-computation was more precise than IPW as seen in the CLD. Finally, as simulation model parameters were less variable (i.e., the secret trial was larger), the CLD was smaller. However, the CLD for the synthesis estimators was still larger than the restriction approaches.

When varying the size of the trial (i.e., $n_0 = 500$), results for bias were similar. The CLD was larger in this scenario, as expected. Again, g-computation had greater precision than IPW for all estimators considered. Further simulation model parameter selections that allowed for uncertainty (i.e., every choice by strict null) had larger CLD than either restriction approach.

\begin{table}[h]
	\caption{Simulation results for $n_1 = 1000, n_0 = 1000$}
	\centering
	\begin{tabular}{llllccc}
		\hline
		\multicolumn{4}{l}{}                           & \multicolumn{1}{l}{Bias} & \multicolumn{1}{l}{CLD} & \multicolumn{1}{l}{95\% CI Coverage} \\ \cline{5-7} 
		\multicolumn{4}{l}{Restrict target population} &                          &                         &                                      \\
		& \multicolumn{3}{l}{G-computation}        & 0.106                    & 0.112                   & 4\%                                  \\
		& \multicolumn{3}{l}{IPW}                  & 0.106                    & 0.159                   & 27\%                                 \\
		\multicolumn{4}{l}{Restrict covariate set}     &                          &                         &                                      \\
		& \multicolumn{3}{l}{G-computation}        & 0.106                    & 0.112                   & 4\%                                  \\
		& \multicolumn{3}{l}{IPW}                  & 0.106                    & 0.158                   & 26\%                                 \\
		\multicolumn{4}{l}{Model synthesis}            &                          &                         &                                      \\
		& \multicolumn{3}{l}{G-computation}        &                          &                         &                                      \\
		&     & \multicolumn{2}{l}{Strict null}    & 0.105                    & 0.111                   & 4\%                                  \\
		&     & \multicolumn{2}{l}{Uncertain null} & 0.078                    & 0.43                    & 100\%                                \\
		&     & \multicolumn{2}{l}{Secret trial $n=2000$}   &                          &                         &                                      \\
		&     &     & Accurate                     & -0.002                   & 0.168                   & 98\%                                 \\
		&     &     & Inaccurate                   & 0.205                    & 0.163                   & 0\%                                  \\
		&     &     & Accurate with covariance     & -0.002                   & 0.151                   & 96\%                                 \\
		&     &     & Reversed                     & 0.000                    & 0.168                   & 97\%                                 \\
		&     &     & Different age distribution   & -0.001                   & 0.164                   & 97\%                                 \\
		&     & \multicolumn{2}{l}{Secret trial $n=4000$}   &                          &                         &                                      \\
		&     &     & Accurate                     & 0.000                    & 0.142                   & 97\%                                 \\
		&     &     & Inaccurate                   & 0.205                    & 0.138                   & 0\%                                  \\
		&     &     & Accurate with covariance     & 0.000                    & 0.131                   & 95\%                                 \\
		&     &     & Reversed                     & 0.001                    & 0.142                   & 97\%                                 \\
		&     &     & Different age distribution   & -0.001                   & 0.139                   & 96\%                                 \\
		& \multicolumn{3}{l}{IPW}                  &                          &                         &                                      \\
		&     & \multicolumn{2}{l}{Strict null}    & 0.106                    & 0.157                   & 28\%                                 \\
		&     & \multicolumn{2}{l}{Uncertain null} & 0.078                    & 0.448                   & 100\%                                \\
		&     & \multicolumn{2}{l}{Secret trial $n=2000$}   &                          &                         &                                      \\
		&     &     & Accurate                     & -0.001                   & 0.201                   & 96\%                                 \\
		&     &     & Inaccurate                   & 0.205                    & 0.195                   & 2\%                                  \\
		&     &     & Accurate with covariance     & 0.000                    & 0.187                   & 95\%                                 \\
		&     &     & Reversed                     & -0.003                   & 0.201                   & 97\%                                 \\
		&     &     & Different age distribution   & 0.001                    & 0.198                   & 96\%                                 \\
		&     & \multicolumn{2}{l}{Secret trial $n=4000$}   &                          &                         &                                      \\
		&     &     & Accurate                     & 0.000                    & 0.179                   & 97\%                                 \\
		&     &     & Inaccurate                   & 0.207                    & 0.174                   & 1\%                                  \\
		&     &     & Accurate with covariance     & 0.000                    & 0.17                    & 95\%                                 \\
		&     &     & Reversed                     & -0.001                   & 0.179                   & 97\%                                 \\
		&     &     & Different age distribution   & -0.001                   & 0.177                   & 96\%                                 \\ \hline
	\end{tabular}
	\floatfoot{CLD: confidence limit difference, CI: confidence interval, IPW: inverse probability weighting. Results are for 2000 iterations.\\
	Bias was defined as mean of the difference between the estimate and true value, where the true value was based on the potential outcomes of 10 million simulated observations.\\
	CLD was defined as the mean of the difference between the upper and lower CI. 95\% CI coverage was defined as the proportion of intervals that contained the truth.}
	\label{atab1}
\end{table}

\begin{table}[h]
	\caption{Simulation results for $n_1 = 1000, n_0 = 500$}
	\centering
	\begin{tabular}{llllccc}
		\hline
		\multicolumn{4}{l}{}                           & Bias   & CLD   & 95\% CI Coverage \\ \cline{5-7} 
		\multicolumn{4}{l}{Restrict target population} &        &       &                  \\
		& \multicolumn{3}{l}{G-computation}        & 0.103  & 0.158 & 28\%             \\
		& \multicolumn{3}{l}{IPW}                  & 0.106  & 0.221 & 53\%             \\
		\multicolumn{4}{l}{Restrict covariate set}     &        &       &                  \\
		& \multicolumn{3}{l}{G-computation}        & 0.103  & 0.158 & 28\%             \\
		& \multicolumn{3}{l}{IPW}                  & 0.106  & 0.223 & 53\%             \\
		\multicolumn{4}{l}{Model synthesis}            &        &       &                  \\
		& \multicolumn{3}{l}{G-computation}        &        &       &                  \\
		&     & \multicolumn{2}{l}{Strict null}    & 0.101  & 0.157 & 29\%             \\
		&     & \multicolumn{2}{l}{Uncertain null} & 0.075  & 0.443 & 100\%            \\
		&     & \multicolumn{2}{l}{Secret trial $n=2000$}   &        &       &                  \\
		&     &     & Accurate                     & -0.005 & 0.198 & 97\%             \\
		&     &     & Inaccurate                   & 0.201  & 0.194 & 2\%              \\
		&     &     & Accurate with covariance     & -0.005 & 0.185 & 95\%             \\
		&     &     & Reversed                     & -0.002 & 0.198 & 96\%             \\
		&     &     & Different age distribution   & -0.003 & 0.195 & 97\%             \\
		&     & \multicolumn{2}{l}{Secret trial $n=4000$}   &        &       &                  \\
		&     &     & Accurate                     & -0.003 & 0.176 & 97\%             \\
		&     &     & Inaccurate                   & 0.202  & 0.174 & 1\%              \\
		&     &     & Accurate with covariance     & -0.003 & 0.168 & 96\%             \\
		&     &     & Reversed                     & -0.002 & 0.177 & 97\%             \\
		&     &     & Different age distribution   & -0.003 & 0.175 & 96\%             \\
		& \multicolumn{3}{l}{IPW}                  &        &       &                  \\
		&     & \multicolumn{2}{l}{Strict null}    & 0.104  & 0.222 & 56\%             \\
		&     & \multicolumn{2}{l}{Uncertain null} & 0.076  & 0.474 & 100\%            \\
		&     & \multicolumn{2}{l}{Secret trial $n=2000$}   &        &       &                  \\
		&     &     & Accurate                     & -0.001 & 0.251 & 96\%             \\
		&     &     & Inaccurate                   & 0.203  & 0.245 & 13\%             \\
		&     &     & Accurate with covariance     & -0.001 & 0.239 & 95\%             \\
		&     &     & Reversed                     & -0.004 & 0.251 & 96\%             \\
		&     &     & Different age distribution   & -0.001 & 0.248 & 97\%             \\
		&     & \multicolumn{2}{l}{Secret trial $n=4000$}   &        &       &                  \\
		&     &     & Accurate                     & -0.001 & 0.234 & 96\%             \\
		&     &     & Inaccurate                   & 0.204  & 0.229 & 8\%              \\
		&     &     & Accurate with covariance     & -0.001 & 0.227 & 95\%             \\
		&     &     & Reversed                     & -0.002 & 0.233 & 96\%             \\
		&     &     & Different age distribution   & -0.001 & 0.232 & 96\%             \\ \hline
	\end{tabular}
	\floatfoot{CLD: confidence limit difference, CI: confidence interval, IPW: inverse probability weighting. Results are for 2000 iterations.\\
		Bias was defined as mean of the difference between the estimate and true value, where the true value was based on the potential outcomes of 10 million simulated observations.\\
		CLD was defined as the mean of the difference between the upper and lower CI. 95\% CI coverage was defined as the proportion of intervals that contained the truth.}
	\label{atab2}
\end{table}

\end{document}